\documentclass[letterpaper]{JHEP3}
\pdfoutput=1

\usepackage[pdftex]{graphicx}
\usepackage{afterpage}
\usepackage{amsmath}

\def\simgt{\mathrel{\lower2.5pt\vbox{\lineskip=0pt\baselineskip=0pt
           \hbox{$>$}\hbox{$\sim$}}}}
\def\simlt{\mathrel{\lower2.5pt\vbox{\lineskip=0pt\baselineskip=0pt
           \hbox{$<$}\hbox{$\sim$}}}}

\title{ Effective AdS/renormalized CFT}

\author{JiJi Fan\\
       Department of Physics, Princeton University, Princeton, NJ, 08540\\
      E-mail: \email{jijifan@princeton.edu}}

\abstract{For an effective AdS theory, we present a simple prescription to compute the renormalization of its dual boundary field theory. In particular, we define anomalous dimension holographically as the dependence of the wave-function renormalization factor on the radial cutoff in the Poincare patch of AdS. With this definition, the anomalous dimensions of both single- and double- trace operators are calculated. Three different dualities are considered with the field theory being CFT, CFT with a double-trace deformation and spontaneously broken CFT\@. For the second dual pair, we compute scaling corrections at the UV and IR fixed points of the RG flow triggered by the double-trace deformation. For the last case, we discuss whether our prescription is sensitive to the AdS interior or equivalently, the IR physics of the dual field theory. }

\newcommand{\beq}{\begin{equation}}
\newcommand{\eeq}{\end{equation}}
\newcommand{\beqs}{\begin{eqnarray}}
\newcommand{\eeqs}{\end{eqnarray}}

\newcommand{\calo}{{\cal{O}}}

\newcommand{\x}{\textbf{x}}
\newcommand{\z}{\textbf{z}}
\newcommand{\0}{\textbf{0}}
\newcommand{\f}{\tilde{f}}
\newcommand{\km}{\textbf{k}}
\newcommand{\pv}{\textbf{p}}

\newcommand{\calk}{{\cal{K}}}

\begin{document}
\section{Introduction}
\label{sec:introduction}
AdS/CFT correspondence~\cite{Maldacena:1997re, Gubser:1998bc, Witten:1998qj} has inspired many phenomenological models, e.g., Randall-Sundrum models~\cite{Randall:1999ee, Randall:1999vf} and their descendants, AdS/QCD models~\cite{Erlich:2005qh, DaRold:2005zs} and most recently, ``bottom-up" AdS/condensed matter models (e.g.,~\cite{Hartnoll:2008vx}, for lectures, see \cite{Herzog:2009xv, McGreevy:2009xe}). The phenomenological models always assume the existence of a general duality between a local bulk effective theory and a (non-supersymmetric and sometimes even non-conformal) field theory away from the infinite $N$ and 't Hooft coupling $g^2N$ limit. Recently, it is proposed by~\cite{Heemskerk:2009pn} that such a duality does exist for any CFT with a few low dimension operators separated by a hierarchy from the dimensions of other operators and a small parameter such as $1/N$ in the large $N$ gauge theory. 

Given the existence of such a duality, it is natural to ask how the CFT gets renormalized by bulk interactions, especially what are the scaling corrections of the operators. In this paper, we will develop a systematic way to compute the anomalous dimensions of both single- and double-trace operators at $\calo(1/N^2)$. To have a well-defined computation in the AdS space, an IR radius cutoff $z_0=\epsilon$ is introduced as the regulator. This way resembles the Wilsonian renormalization with $z_0=\epsilon$ playing the role of the UV cutoff $\Lambda$ in the field theory~\cite{Susskind:1998dq}.  The boundary action at $z_0=\epsilon$ consists of two parts
\beq
S_\epsilon=S_\epsilon^{{\rm{local}}}+S_\epsilon^{{\rm{non-local}}},
\eeq
where $S_\epsilon^{{\rm{local}}}$ consists of all possible (high-dimensional) counter-terms as conformal symmetry is explicitly broken by the radial regulator while $S_\epsilon^{{\rm{non-local}}}$ gives the correlation function of the dual field theory. Notice that these two parts of the action are not independent. The holographic renormalization group (RG) of $S_\epsilon^{{\rm{local}}}$ at the classical level has been discussed within the RS1 setup~\cite{Lewandowski:2002rf, Lewandowski:2004yr}, recently in more general setup with an emphasis on their relation to the the multi-trace flows in the dual field theory in~\cite{Vecchi:2010dd, Heemskerk:2010hk, Faulkner:2010jy}. In this paper, however, we will focus on the renormalization of the correlation function which resides in $S_\epsilon^{{\rm{non-local}}}$. Early investigations on that could be found in~\cite{deBoer:1999xf, deBoer:2000cz}. 

In the simple prescription for the two-point renormalization we proposed, the anomalous dimension measures how wave-function renormalization scales with the radial cutoff $\epsilon$. They contain important information on the whether the operators are radiatively protected and where the effective description breaks down. Notice that anomalous dimensions of double-trace operators have been calculated before by other methods, e.g., through partial wave decomposition of the four-point function of the single-trace operators~\cite{Heemskerk:2009pn, Liu:1998th, D'Hoker:1998mz, D'Hoker:1999pj, D'Hoker:1999jp, Cornalba:2006xm, Cornalba:2007zb} or through perturbations of dilatation operator in the global AdS~\cite{Fitzpatrick:2010zm}. We formulate our calculation in an RG way which allows us to compute in AdS with different boundary conditions. We also take the opportunity to investigate whether the prescription is sensitive to the AdS interior boundary conditions. We find that while it is unclear for the single-trace renormalization, the double-trace renormalization is insusceptible to the IR physics.

The paper is organized as follows. In Sec. 2, we briefly review the correlator function calculations in AdS/CFT and outline the prescription to calculate the anomalous dimensions of CFT operators at $\calo(1/N^2)$. We check it through several toy AdS models with the standard quantization in Sec. 3. In Sec. 4, we proceed to show that our prescription also works for CFT with a double-trace deformation, which corresponds to AdS with mixed boundary conditions. We discuss whether the results are dependent of the AdS interior boundary conditions in Sec. 5. Finally we conclude and point out some open questions in Sec. 6.

\section{Floating radius as the regulator}
\subsection{Correlator from AdS/CFT}
Let us first recall some basic facts about AdS/CFT correspondence. Throughout the paper, we will use the metric of the Euclidean $AdS_{d+1}$ Poincare patch 
\beq
ds^2=\frac{dz_0^2+\sum_{i=1}^d dx_i^2}{z_0^2},
\eeq
where the radial coordinate $z_0$ is restricted to $z_0 \ge \epsilon > 0$ and the AdS radius is set to be 1. For a free scalar field $\phi$ with bulk mass $m$, the solution to the wave function near the boundary $z_0 \to 0$ has the form 
\beqs
\phi (z_0,\x) &=& z^{\Delta_+} [ A(\x)+\calo(z_0^2)] + z^{\Delta_-}[ \phi_0(\x)+\calo(z_0^2)], \\
{\rm{where}} \quad \Delta_\pm&=&\frac{d}{2}\pm \nu, \quad \nu=\sqrt{\frac{d^2}{4}+m^2}.
\eeqs
From now on, bold-faced characters like $\x$ denote the d-vectors like  $\vec{x}$. 
The Breitenlohner-Freedman (BF) bound~\cite{Breitenlohner:1982bm} $m^2\ge -d^2/4$ has to be satisfied for the theory to be stable. For general masses above the BF bound, the scalar theory could be quantized with the Dirichlet boundary condition fixing the boundary value $\phi_0$ of the scalar field $\phi$. The ansatz that relates the boundary CFT to AdS space is that
\beq
\langle exp \int_\epsilon \phi_0 \calo \rangle_{CFT} = Z_\epsilon(\phi_0).
\eeq
The left side of the equation is the generating function of CFT with $\phi_0(\x)$ acting as the source of the dual single-trace scalar operator $\calo$ while the right side is the supergravity partition function on the boundary $Z_\epsilon(\phi_0) \equiv exp(-I_\epsilon(\phi))|_{\phi \to \phi_0}$. With this ansatz, the operator correlator is related to the non-local 1PI boundary action in terms of the source fields $\phi_0$ by a simple rescaling~\cite{PerezVictoria:2001pa}
\beq
\langle {\calo}(\x) {\calo}(\0) \rangle = \lim_{\epsilon \rightarrow 0} \epsilon^{2(d-\Delta)}\langle \phi_0(\x) \phi_0(\0) \rangle_{non-local}^{1 PI}.
\label{eq: 1PI}
\eeq

In the range $-d^2/4 \le m^2 < -d^2/4+1$ or equivalently, $0\le \nu<1$, Dirichlet or Neuman boundary conditions corresponding respectively to fixing $\phi_0$ or $A$ are both possible as both solutions to the wave equations are normalizable. Thus we have two different quantizations, the standard quantization as described above and the alternative quantization where $A(\x)$ serves as the source of the boundary operator $\calo$ with dimension $\Delta_-$. These two quantizations are related by a Legendre transformation in the large $N$ limit \cite{Klebanov:1999tb}. In fact, in this mass range, one could impose a general class of boundary condition
\beq
A(\x)-f \phi_0(\x)=0.
\eeq
In the dual CFT, this corresponds to a double-trace perturbation $\tilde{f}\calo^2$ where $\tilde{f}$ is related to $f$ up to a constant. In the UV, $f , \tilde{f} \to 0$, the CFT is quantized in the alternative way in which $\calo$ has dimension $\Delta_-$. The double-trace deformation is then relevant as $2\Delta_- < d$ and triggers an RG flow which ends in the IR with CFT in the standard quantization and $\calo$'s dimension approaching $\Delta_+$.~\footnote{As an aside, there are several interesting phenomenological proposals based on CFT with double-trace deformations~\cite{Strassler:2003ht, Kaplan:2009kr, Sundrum:2009gv}, some of which require a better understanding of the ``quantum effects" ($\calo(1/N^2)$ effect) in this class of scenario.}

\subsection{Prescription}
\label{sec: prescription}
In this section we outline a simple prescription for computing anomalous dimensions of CFT operators arising from bulk interactions in $AdS_{d+1}$ at $\calo(1/N^2)$. For simplicity, following~\cite{Heemskerk:2009pn, Fitzpatrick:2010zm}, we will consider CFT with only scalar operators. Our examples contain one scalar operator $\calo(\x)$ lying at the bottom of the spectrum, with dimension $\Delta$. This operator is referred to as ``single-trace operator" in analogy to large $N$ gauge theories with adoint representations. Other single-trace operators have much larger dimensions and are decoupled from the low-energy theory. We also assume, according to the conjecture stated at the beginning of the introduction, there exists a small expansion parameter such as $1/N$ in the CFT. At the zeroth order in $1/N$, the primary operators appearing in the $\calo \times \calo$ operator product expansion (OPE) are the ``double-trace operators" 
\beq
\calo_{n,l}(\x) \equiv \calo (\overset\leftrightarrow{\partial}_\mu \overset\leftrightarrow{\partial^\mu})^n (\overset\leftrightarrow{\partial}_{\nu_1} \overset\leftrightarrow{\partial}_{\nu_2}\cdots \overset\leftrightarrow{\partial}_{\nu_l}) \calo(\x), 
\eeq
where parameters $n$ and $l$ denote the twist and the spin of the operator. At the zeroth order in $1/N$, the double-trace operator dimension is $\Delta_{n,l}=2\Delta+2n+l$. 

Once the bulk interactions are turned on, the CFT two-point function would be renormalized. The leading order correction to the bare two-point correlator can be obtained from the non-zero leading order expansion of the boundary 1-PI action
\beq
\epsilon^{2(d-\Delta)}\langle \phi_0(\x) \phi_0(\0)e^{S_{{\rm{bulk}}}(\phi)} \rangle_{non-local}^{1 PI},
\eeq
where integration over the AdS space is involved. As we will see from examples, generally this integration is divergent and a cutoff in the radial direction as $z_0=\epsilon$ is necessary for regulating the integration. 

On the CFT side, we introduce the wavefunction renormalization factor $Z$ which relates renormalized operator ${\calo}_R$ to the bare one ${\calo}$:
\beq
{\calo}_R(\x)=Z{\calo}(\x).
\eeq
The renormalized correlation function
\beq
\langle {\calo}_R(\x){\calo}_R(\0)\rangle= Z^2 \langle {\calo}(\x) {\calo}(\0) \rangle
\eeq
is required to be finite after taking away the cutoff, e.g, $\epsilon \to 0$. Thus the divergent cutoff dependences has to be absorbed in the renormalization factor. The anomalous dimension is then defined as 
\beq
\gamma \equiv -\epsilon \frac {\partial \log{Z}}{\partial \epsilon}.
\label{eq: dimension}
\eeq

It is easy to understand this definition from the CFT point of view. At an interacting fixed point, assuming small anomalous dimension, one could Taylor-expand the two-point function as 
\beq
\langle {\calo}(\x){\calo}(\0) \rangle=\frac{1}{|\x|^{2(\Delta+\gamma)}}=\frac{1}{|\x|^2}{(1-2\gamma \log (|\x|\Lambda))},
\label{eq: powerlaw}
\eeq
where $\Lambda$ is the UV cutoff of the CFT. The $\log \Lambda$ part is to be absorbed into the wavefunction renormalization factor and thus $\gamma = \partial \log{Z}/ \partial \log{\Lambda}$. By the UV/IR duality, the UV cutoff of the CFT is dual to the IR radial cutoff in AdS. For renormalization of the CFT from an interaction in AdS, $\Lambda$ would be replaced by $1/\epsilon$ and then we arrive at Eq.~\ref{eq: dimension}.

\section{Examples}
We now demonstrate how our prescription works by several simple examples with $\nu \equiv \sqrt{d^2/4+m^2} >1$, in which case only the standard quantization is allowed. In the first two examples, we consider bulk mass perturbations. These two examples are trivial in the sense that they could be solved exactly but as we will see, the computations involved serve as the main ingredients for more complicated examples in Sec.~\ref{sec: interaction}.
 
\subsection{Example 1: mass perturbations}
\label{sec: ex1}
\subsubsection{Example 1.1}
We add a tiny mass perturbation $\delta {\cal{V}}_1=1/2\, \delta m^2 \phi^2$ to the bulk field $\phi$. The modified dimension could be solved exactly 
\beqs
\Delta_{\rm{exact}} =  \frac{d}{2} + \sqrt{\left(\frac{d}{2}\right)^2+m^2+ \delta m^2}  =  \Delta_0 + \frac{\delta m^2}{2 \nu}  + {\calo}((\delta m^2)^2), 
\eeqs
where $\Delta_0= d/2  + \nu$ is the dimension of the operator without the mass perturbation. 

Alternatively, one could compute the leading order correction to the boundary correlator
\beq
\langle \phi_0(\x) \phi_0(\0) \int \frac{d z_0d^d z} {z_0^{d+1}} \frac{1}{2}\delta m^2 \phi(z_0,\z)^2  \rangle^{1 PI}_{non-local}\quad.
\label{eq: correction}
\eeq
Among the corrections, the analytic terms give rise to contact terms in position space while the correlator corrections are encoded in the leading non-analytic term. High order terms are unimportant in the limit where the cutoff is removed. Thus the goal is to extract the leading non-analytic term. To achieve that, it is more convenient to work with the mixed momentum-position representation where the field is 
\beqs
\phi(z_0,\km)=\int d^d \z \, e^{i \km \cdot \z}\phi(z_0,\z). \nonumber
\eeqs
For the Dirichlet boundary condition $\phi(\epsilon, \km) = \phi_0(\km)$, the unique solution that is regular for $z_0 \to \infty$ is 
\beqs
\phi(z_0,\km)=\calk(z_0,\km)\phi_0(\km), \nonumber 
\eeqs
with the boundary to bulk propagator  
\beq
\calk(z_0, \km)=\left(\frac{z_0}{\epsilon}\right)^{d/2} \frac{K_\nu(kz_0)}{K_\nu(k\epsilon)}, 
\eeq
where $k=|\km|$ and $K_\nu(kz_0)$ is the modified Bessel function of the second kind. 

In this representation, Eq.~\ref{eq: correction} turns into 
\beqs
&&\langle \phi_0(\x) \phi_0(\0) \int \frac{d z_0d^d z} {z_0^{d+1}} \frac{1}{2}\delta m^2 \phi(z_0,\z)^2  \rangle^{1 PI}_{non local}\nonumber \\
&=& \epsilon^{-d}\,\delta m^2\int _\epsilon\frac{d z_0}{z_0}\int \frac{d^d k}{(2\pi)^d} \left ( \frac{K_\nu(kz_0)}{K_\nu(k\epsilon)} \right)^2 e^{-i \km \cdot \x} \nonumber \\
&=& \cdots-  \epsilon^{2\nu-d}\,2\delta m^2\int _\epsilon\frac{d z_0}{z_0}\int \frac{d^d k}{(2\pi)^d}\left(\frac{k}{2}\right)^{2\nu} \frac{\Gamma(1-\nu)}{\Gamma(1+\nu)}e^{-i \km \cdot \x}  +\cdots \nonumber\\
&=& \cdots+ \epsilon^{2\nu-d}\, \log{\epsilon}\,\frac{2\delta m^2}{\pi^{d/2}}\frac{\Gamma(\Delta)}{\Gamma(\Delta-d/2)} |\x|^{-2\Delta}+\cdots,
\label{eq: mass1}
\eeqs
where in the last two lines we expand the Bessel functions for small arguments $kz, k\epsilon \ll 1$, perform the integral and keep only the leading logarithmic divergent non-analytic term.

Notice that with the bulk-boundary correlator we chose, the two-point function at the classical level is normalized as~\cite{Freedman:1998tz}
\beq
\langle {\calo}(\x){\calo}(\0) \rangle^{(0)}=\frac{2\nu}{\pi^{d/2}}\frac{\Gamma(\Delta)}{\Gamma(\Delta-d/2)} |\x|^{-2\Delta}.
\eeq
To keep two-point correlator finite at order $\delta m^2$, we must have
\beq
Z^2-1=-\frac{\delta m^2}{\nu}\log{\epsilon}+{\rm finite\,\,terms}.
\eeq

By using Eq. \ref{eq: dimension}, we got 
\beq
\gamma=-\epsilon \frac {\partial \log{Z}}{\partial \epsilon}= \frac{\delta m^2}{2\nu},
\label{eq: dimension1}
\eeq
which agrees with the leading order expansion of the exact result for small $\delta m^2$. 

\FIGURE[h]{
\begin{tabular}{ccc}
\includegraphics[scale=0.5]{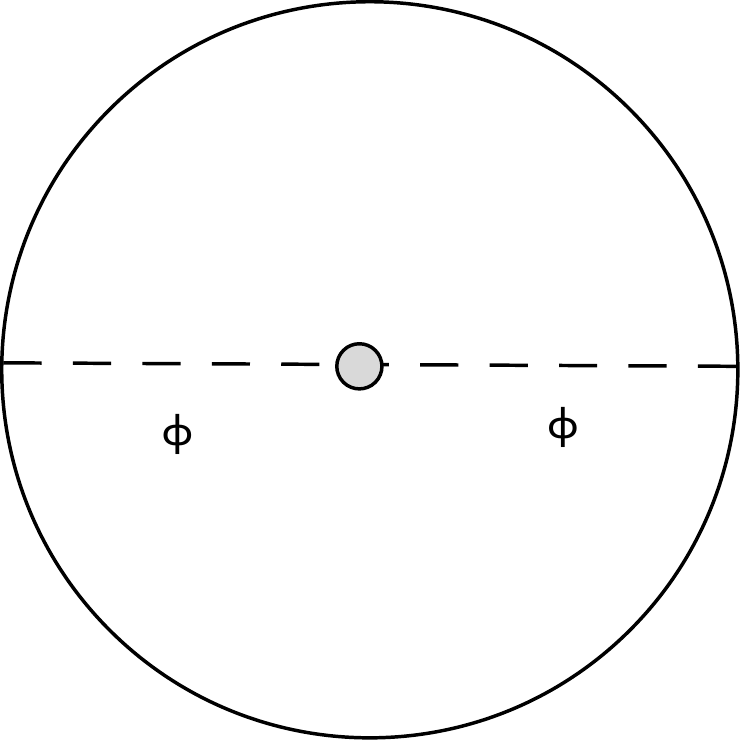}& 
\includegraphics[scale=0.5]{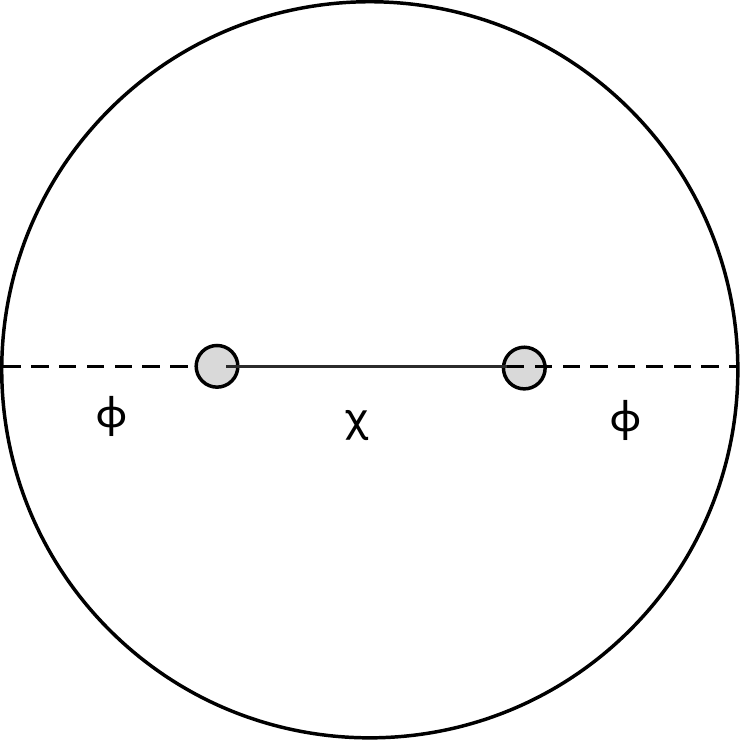} &
\end{tabular}
\caption{Witten diagrams for the models with mass perturbations. The blobs represent insertions of mass perturbation. Left: example 1.1; right: example 1.2.}
\label{fig: mass}
}

\subsubsection{Example 1.2}
\label{sec: ex1.2}
In the second example, we consider a slightly more complicated mass structure in a scalar theory of two scalars $\phi, \chi$ with the bulk potential density
\beq
{\cal{V}}= -\frac{1}{2}m_1^2 \phi^2-\frac{1}{2}m_2^2 \chi^2-\delta m^2 \phi \chi, \quad \delta m^2 \ll m_1^2 \ne m_2^2 \, ,
\eeq

One could diagonalize the mass matrice and expand the exact solution in terms of small mixing $\delta m^2$. For instance, for $\phi$, 
\beq
\gamma_{1} = \frac{(\delta m ^2)^2}{m_1^2-m_2^2} \frac{1}{2\nu_{1}}. 
\label{eq: dimension2}
\eeq
Again one could also calculate the leading correction to the boundary two-point function from two mass insertions in the bulk (the right plot in Fig.~\ref{fig: mass})
\beq
\langle \phi_0(x) \phi_0(0) \rangle^{(1)} = (\delta m^2)^2 \int \frac{d z_0}{z_0^{d+1}}\frac{d w_0}{w_0^{d+1}}\int \frac{d^d \km}{(2\pi)^d} e^{-i \km \cdot \x} \calk(z_0,\km) G(z_0,w_0; \km) \calk(w_0, \km),
\label{eq: mass2}
\eeq
where for the Dirichlet boundary condition, the bulk-to-bulk propagator $G(z_0,w_0; \km)$ reads~\cite{Muck:1998rr}
\beq
G(z_0,w_0;\km)= z_0^{d/2}w_0^{d/2}\left( \left(I_\nu(kw_0)K_\nu(kz_0)-K_\nu(kw_0)K_\nu(kz_0) \frac{I_\nu(k\epsilon)}{K_\nu(k\epsilon)}\right)\theta(z_0-w_0) + z_0 \leftrightarrow w_0 \right)
\label{eq: bulkpro}
\eeq
with $I_\nu(z)$ the modified Bessel function of the first kind.

Plugging it into Eq. \ref{eq: mass2}, one finds that only one term in the bulk-to-bulk propagator contributes
\beq
\langle \phi_0(x) \phi_0(0) \rangle^{(1)}= \epsilon^{-d}\,2(\delta m^2)^2\int_\epsilon \frac{d z_0}{z_0}  \frac{K_{\nu_1}(k z_0)}{K_{\nu_1}(k\epsilon)}K_{\nu_2}(k z_0) \int^{z_0}_\epsilon \frac{dw_0}{w_0} \frac{K_{\nu_1}(k w_0)}{K_{\nu_1}(k\epsilon)}I_{\nu_2}(k w_0) ,
\eeq
where the factor 2 takes into account of the fact that $w_0>z_0$ contributes exactly the same as $z_0>w_0$. Expanding the Bessel functions in terms of small arguments and keeping track of only the leading non-local term, we arrive at the same answer in Eq.~\ref{eq: dimension2}. 

As a consistency check, if one field is much heavier than the other one, e.g., $m_2^2 \ll m_1^2$, $\chi$ could be integrated out.  Then example 1.2 would turn into example 1.1 with the matching condition
\beq
\delta{\cal{V}}_1= \frac{(\delta m^2)^2}{2 m_2^2} \phi^2.
\eeq
The final results of the two examples agree in this limit. 

It is easy to see that Eq.~\ref{eq: dimension2} has a pole at $\Delta_1=\Delta_2$ and close to this value there is a resonance-like enhancement of $|\gamma|$. Fixing $\Delta_2$, when $\Delta_1$ increases from a small value compared to $\Delta_2$, the anomalous dimension grows. Once it passes the critical value $\Delta_2$, the growth would be cut off in the ``full theory" by ``integrating in" the heavy primary $\chi$. This is a toy version of the resonance behavior for double-trace renormalization from exchange interactions which we will discuss in Sec. 3.2.3.

\subsection{Example 2: contact interaction}
\label{sec: interaction}
In this section, we will consider both single- and double-trace renormalizations from bulk contact interactions, e.g., a $\mu\phi^4/4!$ interaction. As one will see, the main parts of both computations have already been performed in our first toy example with a bulk mass shift. We will also consider double-trace renormalization from exchange of an additional heavy scalar field.

\subsubsection{Single-trace  renormalization}
\FIGURE[h]{
\includegraphics[scale=0.5]{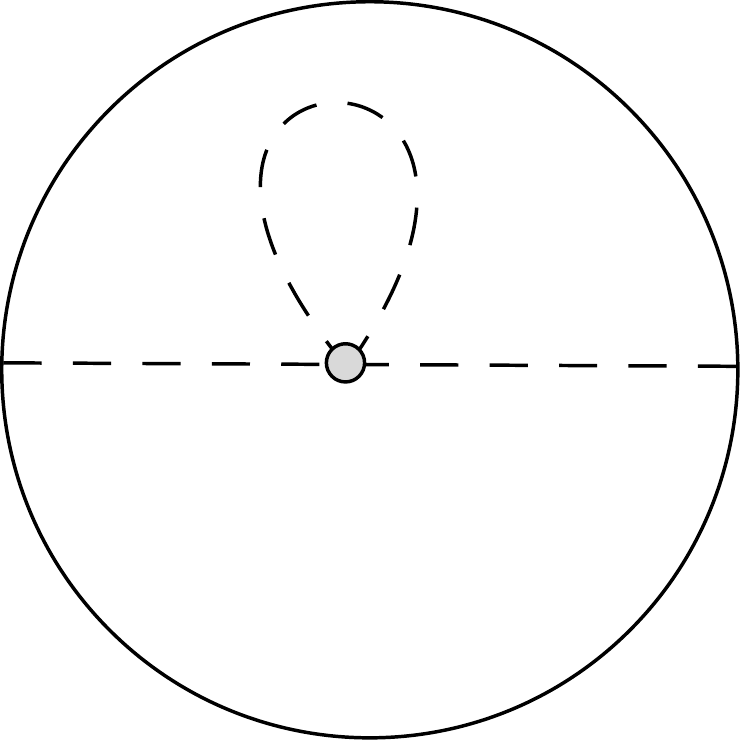} 
\caption{Witten diagram for the leading correction to the single-trace two-point function from the $\phi^4$ contact interaction (denoted by the blob). }
\label{fig: phi4_1}
}
After turning on a $\mu\phi^4/4!$ interaction in the bulk, the single-trace boundary-boundary correlator is renormalized by a one-loop diagram in AdS (Fig.~\ref{fig: phi4_1})
\beqs
&&\langle \phi_0(\x) \phi_0(\0) \int  \frac{d z_0d^d z}{z_0^{d+1}} \frac{\mu}{4!}\phi(z_0,\z)^4  \rangle^{1 PI}_{non-local} \nonumber \\
&=&\frac{\mu}{2} \epsilon^{-d}\int _\epsilon\frac{d z_0}{z_0} \int \frac{d^d k}{(2\pi)^d} \left ( \frac{K_\nu(kz_0)}{K_\nu(k\epsilon)} \right)^2 e^{-i \km\cdot \x}\int \frac{d^d p}{(2\pi)^d} G(z_0, z_0;\pv), \nonumber \\
\eeqs
where $G(z_0, z_0; \pv)$ is the bulk-to-bulk propagator with two coincident external points.

The computation looks almost identical to Eq.~\ref{eq: mass1} after replacing the mass shift $\delta m^2$ by $\mu/2\int d^dp \, G(z_0,z_0;\pv)$. For $\phi^4$ interaction, however, this momentum integration is divergent. As the ultra-violet divergences are local and insensitive to the space-time curvature, the power countings of divergence should be the same in both the flat and warped space. This can be confirmed explicitly by inspecting the asymptotic form of the AdS bulk-to-bulk propagator:
\beqs
{\rm{Flat \,space:}} \quad \delta m^2 &\sim & \int d^{d+1} p \, \frac{1}{p^2} \sim \Lambda^{d-1}, \nonumber \\
{\rm{AdS \, space:}} \quad \delta m^2 &\sim & \int d^{d} p\, G(z_0, z_0; \pv) \sim \int d^d p\, I_\nu(pz_0) K_\nu(pz_0) \sim \int d^d p\, \frac{1}{p}\sim \Lambda^{d-1}. \nonumber
\eeqs

One plausible simple regulator is a position-dependent momentum cutoff $p_{{\rm{cutoff}}} =\alpha/z_0$ with $\alpha$ a dimensionless parameter. This regulator originates from the well-known fact that the effective cutoff scales as the inverse of the radial position $z_0$. In this scheme, the anomalous dimension of single-trace operator is
\beqs
 \gamma= \frac{\delta m^2}{2\nu}, \nonumber 
\eeqs
where $\delta m^2=\mu I(\alpha)$ and
\beqs
 I(\alpha)&=&\frac{\alpha^d}{4(4\pi)^{d/2}\nu\Gamma(d/2+1)}({}_2F_3(\frac{1}{2},\frac{d}{2};1+\frac{d}{2},1-\nu,1+\nu;\alpha^2) \nonumber \\
&& -\alpha^{2\nu}\frac{d\,\Gamma(\frac{1}{2}+\nu)\Gamma(1-\nu)}{\sqrt{\pi}(d+2\nu)\Gamma(1+2\nu)}{}_2F_3(\frac{1}{2}+\nu,\frac{d}{2}+\nu;1+\nu,1+\frac{d}{2}+\nu,1+2\nu;\alpha^2)).\nonumber \\
\label{eq: single ano}
\eeqs

The momentum-dependent cutoff is not the unique choice of regularizing the momentum integration. Other schemes such as the Pauli-Villas regularization could lead to different answers. This is not surprising as the single-trace operator is not radiatively protected in an effective theory such as $\phi^4$ theory. The scheme-dependences of its anomalous dimensions reflects its sensitivity to the UV completion of the effective theory.

\subsubsection{Double-trace renormalization}
Analogous to single-trace operator, we assume $\phi_0^2(x)$ sources double-trace operators \cite{Chalmers:1999gc} and by a rescaling, the correction of the bare operator correlation is related to the correction to the boundary 1PI action from the bulk interaction as
\beq
\langle {\calo_{n,l}}(\x) {\calo_{n,l}}(\0) \rangle^{(1)} = \lim_{\epsilon \rightarrow 0} \epsilon^{2(d-\Delta_{n,l})}\langle \phi_0^2(\x) \phi_0^2(\0) \cdots\rangle_{non-local}^{1 PI},
\eeq
where the dots denote bulk interaction insertions. The expression should be understood as that for each $\calo_{n,l}$, one extracts from $\langle \phi_0^2(x) \phi_0^2(0) \cdots\rangle_{non-local}^{1 PI}$ the term with the corresponding scaling $\epsilon^{2(\Delta_{n,l}-d)}$.
Thus for the $\phi^4$ interaction, one needs to compute
\beqs
\langle \phi_0^2(\x) \phi_0^2(\0) \int  \frac{d z_0d^d z}{z_0^{d+1}} \frac{\mu}{4!}\phi(z_0,\z)^4  \rangle^{1 PI}_{non-local}. 
\label{eq: phi4d}
\eeqs

Apparently the computations of double-trace dimension corrections look quite different from previous ones. However, we would apply a procedure analogous to the conformal block decomposition of CFT correlators \cite{Fitzpatrick:2010zm}, which would reduce the computation to that in example 1.1. Consider the two-point function between two bulk points $\langle \phi^2(z_0, \x) \phi^2(w_0, \0) \rangle$. It can be easily computed from Wick contractions as a product of two single-particle propagators 
\beqs
\langle \phi^2(z_0, \x) \phi^2(w_0, \0) \rangle = 2 G(z_0, w_0; \x, \0)^2, \nonumber 
\eeqs
where $G(z_0,w_0;\x,\0)=\int d^d k/(2\pi)^d e^{-i \km \cdot \x} G(z_0,w_0;\km)$ is the bulk propagator in the position space. 
Instead of using this directly to compute \ref{eq: phi4d}, we would decompose it as a sum of  weighted ``two-particle propagator" $G_{n,0}(z_0,w_0;\x,\0)$
\beqs
\langle \phi^2(z_0, \x) \phi^2(w_0, \0) \rangle=\sum_n \frac{G_{n,0}(z_0,w_0;\x,\0)}{N_n^2}, 
\label{eq: normalization}
\eeqs
where $N_n$ is the normalization factor depending on the twist $n$. $G_{n,0}(z_0,w_0;\x,\0)$ assumes the same form of the single-particle propagator except replacing $\nu$ with $\nu_n \equiv \Delta_{n,0} - d/2$. This could be understood as the conformal symmetry determines all the propagators, whether it is single- or multi- particle's, up to a normalization. As $\phi^2$ only creates and annihilates states without spin, this decomposition is independent of spin.  Details of the calculation for the normalization factors $N_n^2$ could be found in the Appendix. Below we just quote the numbers for $d=2$ and $d=4$ 
\beqs
d&=&2 \quad N_n^{-2}=\frac{1}{\pi}\nonumber \\
d&=&4 \quad N_n^{-2}=\frac{1}{\pi^2}\frac{(n+1)(\Delta+n-1)^2(2\Delta+n-3)}{(2\Delta+2n-1)(2\Delta+2n-3)} 
\label{eq: norm}
\eeqs
Similarly for the bulk-to-boundary propagator, it should follow the same decomposition
\beqs
\langle \phi_0^2(\x) \phi^2(z_0,\z) \rangle = \sum_{n} \frac{\calk_{\nu_n}(z_0; \x, \z)}{N_n^2}, 
\label{eq: normalization2}
\eeqs
with 
\beqs
\calk_{\nu_n}(z_0; \x, \0) \equiv \int \frac{d^d \km}{(2\pi)^d} e^{-i \km \cdot \x} \calk_{\nu_n}(z_0, \km) =\int \frac{d^d \km}{(2\pi)^d} e^{-i \km \cdot \x} \left(\frac{z_0}{\epsilon}\right)^d \frac{K_{\nu_n} (kz_0)}{K_{\nu_n} (k\epsilon)}.  \nonumber
\eeqs

After the decomposition, for each individual double-trace operator, the calculation of the double-trace operator anomalous dimension proceeds similarly as that of example 1.1 with the identification $\delta m^2=\mu/4$ after taking into account of the symmetry factor. For even dimensions $d=2,4$, we find
\beqs
d&=&2 \quad \gamma(n)=\frac{\mu}{8\pi}\frac{1}{2\Delta+2n-1}, \nonumber \\
d&=&4 \quad \gamma(n)=\frac{\mu}{16\pi^2}\frac{(n+1)(\Delta+n-1)(2\Delta+n-3)}{(2\Delta+2n-1)(2\Delta+2n-3)},
\eeqs
which agree with \cite{Heemskerk:2009pn, Fitzpatrick:2010zm}. 

\FIGURE[h]{
\begin{tabular}{ccc}
\includegraphics[scale=0.5]{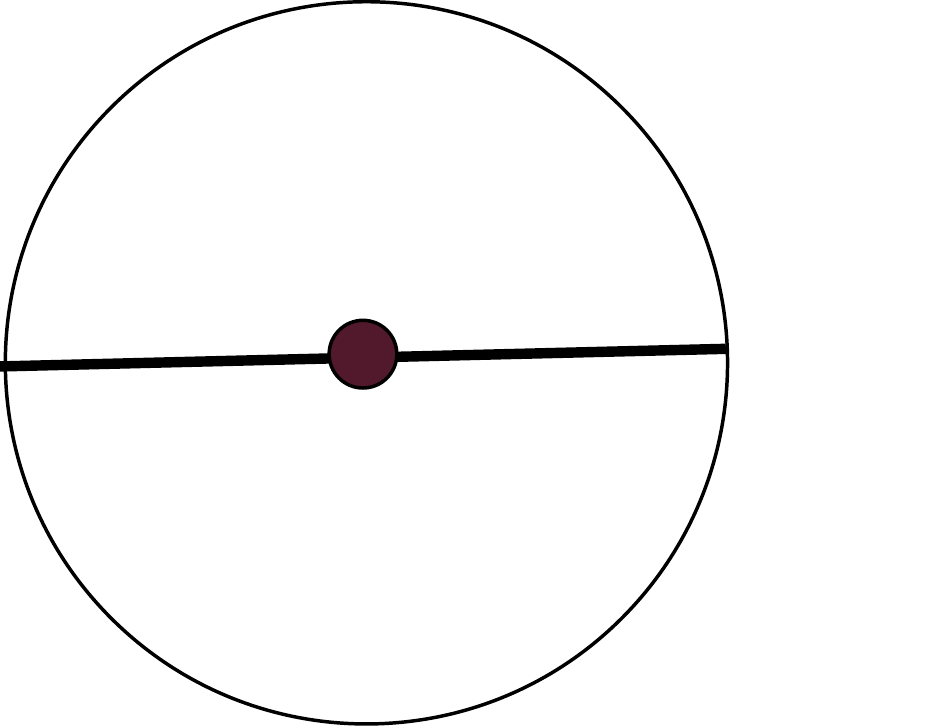} &
\includegraphics[scale=0.6]{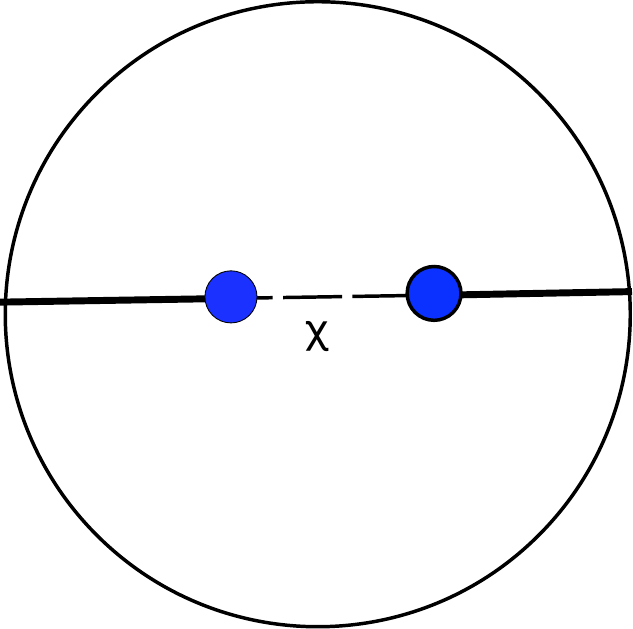}&
\end{tabular}
\caption{Witten diagrams for corrections to double-trace two-point correlator from $\phi^4$ (left plot) and $\phi^2\chi$ (right plot). The blobs denote insertions of the bulk interaction. The solid lines represent the ``two-particle propagator" defined in the text.}
\label{fig: phi4}
}

Traditionally, one needs to extract from the four-point functions of single-trace operators $\langle \calo(x_1)\calo(x_2)\calo(x_3)\calo(x_4) \rangle$ the leading logarithmic singularity of the cross ratios and then project it onto the partial wave associated with one particular double-trace operator to obtain its anomalous dimension~\cite{Liu:1998th, D'Hoker:1998mz, D'Hoker:1999pj, D'Hoker:1999jp, Cornalba:2006xm, Cornalba:2007zb}. Our method, to an extent, is to reverse the two steps of extraction of the leading logarithmic divergence and partial wave decomposition. Another method~\cite{Fitzpatrick:2010zm} performs the calculation in the global coordinates, where $\gamma(n)$ is identified as correction to the eigenvalue of the dilatation operator arising from bulk perturbations. 

One interesting property of $\gamma(n)$ is that~\cite{Heemskerk:2009pn, Fitzpatrick:2010zm}, for large $n$, $\gamma(n) \sim n^{d-3}$. More generally, for a bulk contact interaction $\calo/\Lambda^p$, the anomalous dimension of the double-trace operator with large $n$ always grows as $\gamma(n) \sim n^p$. This could be understood in analogy to the amplitude of the scattering mediated by high-dimensional operators in the ordinary effective field theory with the following identifications
\beqs
\gamma(n) &\leftrightarrow& {\cal{A}}, \nonumber \\
n &\leftrightarrow& E. \nonumber
\eeqs
As is well known, the scattering amplitude generated by an operator $\calo/\Lambda^p$ grows as ${\cal{A}} \sim E^p$. For example, Euler-Langrange operator $F^4/m_e^4$ leads to photon-photon scattering with amplitude ${\cal{A}} \sim E^4$. As energy grows, the amplitude grows. When $E \sim m_e$, the effective theory description breaks down and new degrees of freedom must be ``integrated in" to unitarize the theory. Similarly, for the CFT with heavy enough two- particle states $n \gg 1$, new degrees of freedom must be included on the AdS side to keep the duality valid at the perturbative level. 

\subsubsection{Integrating out heavy state}
Before ending the section of examples, we will comment on the double-trace renormalization from exchange of a heavy scalar in $AdS$. To be concrete, the bulk interaction is 
\beq
 \lambda \int d^d x dz_0 \sqrt{-g}\, \phi^2 \chi (x), \nonumber
\eeq
where $\chi(x)$ is a massive scalar in $AdS_{d+1}$. For $d<6$ this interaction is renormalizable. In the limit $m_\chi \gg m_\phi$, one can integrate out $\chi$ to obtain an effective theory with $\phi^4$ contact term. Below we will compute the contributions to the anomalous dimensions of $\phi$ double-trace operators from $\phi^2\chi$ interaction directly. For simplicity, we would focus on scalar exchange in the s-channel, which only contributes to the spinless anomalous dimension $\gamma(n)$. 

After implementing the conformal partial wave decomposition as in Eq.~\ref{eq: normalization}, \ref{eq: normalization2}, the s-channel computations are reduced to that in Example 1.2 (compare the right plots in Fig. 1 and  Fig. 3). After replacing $\nu_1$ by $\nu_n=2\Delta+2n-d/2$, $\nu_2$ by $\nu_\chi$ and $\delta m^2$ by $\lambda$ in Eq.~\ref{eq: dimension2}, we got \footnote{Our result looks different in formula from that in~\cite{Fitzpatrick:2010zm} but agrees with result derived using method in \cite{Fitzpatrick:2011hh}. The two results could be proven to be the same numerically~\cite{Fitzpatrick:2011}.}
\beqs
\gamma(n) &=&\frac{\lambda^2}{N_n^2} \frac{1}{2(\nu_n^2-\nu_\chi^2)\nu_n}, \\
d=2\quad \gamma(n)&=& \frac{\lambda^2}{2\pi}\frac{1}{(2\Delta+2n+\Delta_\chi-2)(2\Delta+2n-\Delta_\chi)(2\Delta+2n-1)},\nonumber \\
d=4 \quad \gamma(n)&=& \frac{\lambda^2}{2\pi^2}\frac{(n+1)(\Delta+n-1)(2\Delta+n-3)}{(2\Delta+2n+\Delta_\chi-4)(2\Delta+2n-\Delta_\chi)(2\Delta+2n-1)(2\Delta+2n-3)}. \nonumber\\
\eeqs

It is easy to see that all expressions have a pole at $\Delta_n\equiv2\Delta+2n=\Delta_\chi$ and thus a resonance enhancement of $\gamma(n)$ around the pole. At this pole, one needs to go to high order in perturbation theory where the resonance of $\gamma(n)$ would be smoothed out and has a finite width, due to the fact that $\chi$ has a finite lifetime in $AdS$. Passing this pole, the growth is cut off by integrating in $\chi$ and  $\gamma(n)$ will start to decrease. 

\section{CFT with a double-trace deformation}
As we mentioned briefly in Sec 2, for a tachyon field in the bulk with mass in the range $-d^2/4 \le m^2 < -d^2/4+1$, one could impose a general boundary condition corresponding to the presence of a double-trace deformation of the boundary CFT ${\cal{L}}= {\cal{L}}_{CFT}+\frac{f}{2}\calo^2$. Such an operator can lead to an RG flow between two different CFT's related to each other by a Legendre transformation in the large $N$ limit. In this section we will demonstrate how our prescription works in this case. Particularly, we will construct an interpolating function which produces the correct operator scaling dimensions at the UV and IR fixed points.  

As pointed out first in~\cite{Witten:2001ua}, the double-trace deformation corresponds to a mixed Neumann / Dirichlet boundary condition at the boundary $z_0=\epsilon$
\beq
\f \phi(\epsilon,\km) +\epsilon\partial_{z_0} \phi(z_0,\km) |_{z_0=\epsilon}= \phi_0(\km).
\eeq
The parameter $\f$ is related to the double-trace coupling $f$ as ~\cite{Hartman:2006dy}
\beq
\f=-\Delta_- - f \epsilon^{2\nu}\left(2\pi^{d/2}\frac{\Gamma(1-\nu)}{\Gamma(\Delta_-)}\right).
\eeq
The bulk-to-boundary propagator in this case turns out to be 
\beqs
\psi(kz_0)&\equiv& z_0^{d/2} K_\nu(kz_0), \nonumber \\
K(\km,z_0)&=&\frac{\psi(kz_0)}{\f\psi(k\epsilon)+\epsilon\partial_{z_0} \phi(z_0,\km) |_{z_0=\epsilon}}.
\eeqs
For a free theory, the two-point correlator parametrized by $f$, in the momentum space is 
\beqs
\langle \calo(\km) \calo(\km^\prime)\rangle_f^{(0)}&=& -\epsilon^{-d} \delta^{d}(\km+\km^\prime) \frac{\psi(k\epsilon)}{\f\psi(k\epsilon)+\epsilon\partial \psi(k\epsilon)} \nonumber \\
&=& -\epsilon^{-d} \delta^{d}(\km+\km^\prime) \frac{1}{- f \epsilon^{2\nu}\left(2\pi^{d/2}\frac{\Gamma(1-\nu)}{\Gamma(\Delta_-)}\right)+\epsilon^{2\nu}(\frac{k}{2})^{2\nu}\frac{\Gamma(-\nu)}{\Gamma(\nu)}(2\nu)}.
\label{eq: 2pt-int}
\eeqs
Starting at the UV with $f=0$, the ${\rm{CFT}}_{\rm{UV}}$ is in the alternative quantization as one can see from Eq. \ref{eq: 2pt-int}. Flowing down to the IR, $f$ grows as double-trace perturbation is relevant. As $f \to \infty$ in the deep IR, we recover the Dirichlet boundary condition and the two-point correlator is that in the standard quantization up to some overall constant normalizations. 

For the mass perturbation $\delta m^2 \phi^2/2$, the correction to the boundary correlator is 
\beqs
&&\langle \calo(\km) \calo(\km^\prime)\rangle_f^{(1)}=- \delta^{d}(\km+\km^\prime)\delta m^2\int_\epsilon \frac{dz_0}{z_0}\left( \frac{\psi(kz_0)}{\f\psi(k\epsilon)+\epsilon\partial \psi(k\epsilon)} \right)^2\nonumber \\
&&=-\epsilon^{-d}\delta^{d}(\km+\km^\prime)\delta m^2\int_\epsilon \frac{dz_0}{z_0}\left(\frac{2^{\nu-1}\Gamma(\nu)(kz_0)^{-\nu}+2^{-\nu-1}\Gamma(-\nu)(kz_0)^\nu}{- f \epsilon^{\nu}\pi^{d/2}2^{\nu}\frac{\Gamma(1-\nu)\Gamma(\nu)}{\Gamma(\Delta_-)}k^{-\nu}-2^{-\nu}\Gamma(1-\nu)(k\epsilon)^\nu}\right)^2.
\eeqs
Following the same procedure in Sec. 3, at the two ends of the RG flow, we got
\beqs
&&{\rm{UV}}\quad f \to 0\quad \gamma_{{\rm{UV}}}=-\frac{\delta m^2}{2\nu}, \nonumber \\
&&{\rm{IR}}\quad f \to \infty\quad \gamma_{{\rm{IR}}}=\frac{\delta m^2}{2\nu}. \nonumber 
\eeqs
Notice that the anomalous dimension in the UV and IR have opposite signs as one would expect from expansion of the exact result $\Delta_{\pm} = d/2\pm \nu$. 

Computations of other examples proceed similarly. Before writing down the results, we would like to make a few comments on the bulk-to-bulk propagator involved in the computations. As the conformal invariance is explicitly broken by the RG flow, the bulk-to-bulk propagator cannot be $SO(d,2)$ invariant. One could check that the $SO(d,2)$ invariant propagator $G(z_0,w_0; \x, \0; t) = t\, G_{\Delta_+} +(1-t) G_{\Delta_-}$ with $t \in [0,1]$ could not satisfy the mixed boundary condition for arbitrary $t$ except for $t=0$ and $t=1$, corresponding to the two fixed points. However, one could construct an $SO(d,2)$ noninvariant propagator parametrized by $f$ in the mixed representation~\cite{Gubser:2002zh}, which reproduces $G_{\Delta_-}(G_{\Delta_+})$ when $f=0 (f\to\infty)$. Then for the contact interactions, the anomalous dimensions of both single- and double-trace operators at the IR fixed point are the same  as computed in the previous section. At the UV fixed point, $\gamma_{{\rm{UV}}}$ is identical to $\gamma_{{\rm{IR}}}$ with the replacement of $\Delta_+$ by $\Delta_-$, or equivalently, $\nu$ by $-\nu$. 

Notice that for the single-trace renormalization from bulk interaction such as $\phi^4$, $\gamma_{{\rm{UV}}}+\gamma_{{\rm{IR}}}$ is finite as the UV divergences of loop momentum integration cancel each other. Thus $\gamma_{{\rm{UV}}}+\gamma_{{\rm{IR}}}$ is scheme independent. This is similar to computations of the vacuum energy at one loop\cite{Gubser:2002zh, Gubser:2002vv}. While the one-loop self energy diverges, its change between the two fixed points corresponding to a change in the central charge of dual field theory, is finite. At the classical level, $\Delta_{{\rm{UV}}} +\Delta_{{\rm{IR}}}=d$. However,  at $\calo(1/N^2)$, $\Delta_{{\rm{UV}}} +\Delta_{{\rm{IR}}}$ deviate from $d$ in general. 

One caveat we did not explore here is that at $\calo(1/N^2)$, the geometry would deviate from AdS, as in the one-loop diagram where the $SO(d,2)$-noninvariant propagator closes upon itself would give rise to an effective potential varying over spacetime. The inclusion of the loop-induced back-reaction is important but lies beyond this paper. \footnote{Appendix A of Ref.~\cite{Kaplan:2009kr} offers one calculation in a scalar field theory without interactions on how the geometry deviation feeds into the operator scaling dimensions.}

\section{Spontaneous broken CFT - Wilsonian scheme or not?}
So far we have assumed one particular IR boundary condition, the finiteness of the wave function solution at $z \to \infty$, when constructing the propagators and computing the anomalous dimensions. One might worry that whether the anomalous dimensions obtained this way is not what one would get from a Wilson scheme but rather some IR- dependent renormalization scheme~\cite{deBoer:1999xf, deBoer:2000cz}. To test whether our prescription is sensitive to the IR physics, we would explore a different setup with an IR brane. 

When the AdS space is truncated at finite radius $z=L$, it is dual to sponteneous broken CFT, of which the two-point function doesn't need to follow the simple power scaling behavior in the IR. More specifically, at small momentum or large separation $x$, there could be poles appearing in the two point functions corresponding to discrete tower of KK modes. In general, one would expect Eq. \ref{eq: powerlaw} to be modified and correspondingly our prescription to be modified in order to subtract CFT anomalous dimensions that are independent of the IR condition. However, we find that for bulk mass perturbations, the dimension corrections are indeed insensitive to the IR physics following our prescription. 

Specifically, we impose a Dirichlet boundary condition at $z=L$. Then one could parametrize the boundary-to-bulk propagator as 
\beq
\calk(z_0, \km)= \left(\frac{z_0}{\epsilon}\right)^{d/2} \frac{K_\nu(kz_0)+a I_\nu(kz_0)}{K_\nu(k\epsilon)+a I_\nu(k\epsilon)},
\eeq
with ${\rm lim}_{z_0\to\epsilon} \calk(z_0,\km)=1$. Parameter $a$ is fixed by the IR boundary conditions whose specific value is irrelevant here. 
For example 1.1, with this modified propagator, Eq. \ref{eq: mass1} becomes
\beqs
\langle\calo(\x)\calo(\0)  \rangle^{(1)}=\log{\epsilon}\,\frac{2\delta m^2}{\pi^{d/2}}\frac{\Gamma(\Delta)}{\Gamma(\Delta-d/2)}\left(1+2a\frac{1}{\Gamma(1+\nu)\Gamma(-\nu)}\right) |\x|^{-2\Delta},
\label{eq: modification}
\eeqs
which now contains an IR dependent factor. But at the classical level, the normalization of the two point function also changes 
\beqs
\langle {\calo}(\x){\calo}(\0) \rangle^{(0)}=\frac{2\nu}{\pi^{d/2}}\frac{\Gamma(\Delta)}{\Gamma(\Delta-d/2)}\left(1+2a\frac{1}{\Gamma(1+\nu)\Gamma(-\nu)}\right) |\x|^{-2\Delta}.
\eeqs
 The modification of the normalization exactly cancels the $a$ dependent part in Eq. \ref{eq: modification}. Thus the single-trace anomalous dimension $\gamma$ remains unchanged for different interior boundary conditions. 

For example 1.2, the final answer is also independent of the IR boundary conditions. Compared to example 1.1, besides the bulk-to-boundary propagator, the bulk-to-bulk propagator also varies with the IR conditions. For the Dirichlet boundary condition at $z=L$, the bulk-to-bulk propagator is
\beqs
&&G(z_0,w_0;\pv)=z_0^{d/2}w_0^{d/2} \nonumber \\
&&\left(\frac{(-K_\nu(p\epsilon)I_\nu(pw_0)+I_\nu(p\epsilon)K_\nu(pw_0))(-K_\nu(pL)I_\nu(pz_0)+I_\nu(pL)K_\nu(pz_0))}{-K_\nu(pL)I_\nu(p\epsilon)+I_\nu(pL)K_\nu(p\epsilon)}\theta(z_0-w_0)+z_0 \leftrightarrow w_0\right), \nonumber \\
\eeqs
Plugging it into Eq. \ref{eq: mass2}, one finds that indeed all the IR dependences cancel out, leaving the final answer unchanged.

Subtleties arise for anomalous dimensions from contact interactions. For the single-trace renormalization, the divergence counting is determined purely by short-distance physics and thus is independent of the CFT vacuum. However, the finite part of the loop momentum integration does depend on the explicit form of the bulk-to-bulk propagator.  Whether there is a renormalization scheme in which the IR properties factor out of the computation is beyond our knowledge and we will not pursue it here further. 

However, after the partial-wave decomposition in Eq.~\ref{eq: normalization}, the double-trace renormalization is reduced to renormalization induced by a mass perturbation. Thus it is insensitive to the AdS interior!

\section{Conclusion and open questions}
In this article, we present a simple holographic definition of anomalous dimensions arising from bulk interactions in the dual AdS effective field theory. We checked it in toy examples with mass perturbations and contact/exchange interactions. Three classes of dualities are considered, with the field theory being CFT, RG flows between two fixed points and spontaneously broken CFT. For the last class, we identify cases where our definition is independent of the IR boundary conditions, as one expects a true Wilsonian renormalization scheme would achieve.

There are still many questions and directions that might worth exploring. 
\begin{itemize}
\item{In the phenomenological 5D models, it is implicitly assumed that the compact internal dimensions (e.g., $S^5$ in $AdS_5 \times S^5$) are not important. However, in the Maldacena conjecture, the radius of $S^5$ is of order the AdS radius. In that case, instead of having a 5D effective theory description, we would have to deal with a full 10D theory\footnote{We thank Matt Strassler for emphasizing this point.}.  Recently a particular class of string models~\cite{Polchinski:2009ch} has been constructed where the internal dimension size could be parametrically smaller than the AdS curvature. It is still unclear whether pure AdS effective theories as the low-energy descriptions are generic or not and whether they would always have UV completions. For phenomenological implications of (non-decoupled) internal dimensions, see~\cite{Reece:2010xj}.}

\item{The single-trace operators' dimension could be radiatively unstable in an effective AdS theory. This is true not only for the toy model with $\phi^4$ interaction but also for more realistic models (e.g., 5D RS flavor models) with bulk fermions and Yukawa/gauge interactions. One might worry about the naturalness and predictability of these phenomenological models.}

\item{It would be interesting to compute renormalization of boundary field theory in the gauge/gravity system. In our computations, we neglected the scalar potential's back-reaction to the geometry and assumed that such a non-supersymmetric AdS effective theory has no instability. Careful treatments of these issues could be important.}

\item{As discussed in the introduction, there is a rising interest in the relation between the holographic flows of the boundary local couplings and the multi-trace flows in the CFT~\cite{Vecchi:2010dd, Heemskerk:2010hk, Faulkner:2010jy}. Their duality is confirmed at the classical level and it might be interesting to know how to compute the corrections to flows on both sides at the quantum level (or at $\calo(1/N^2)$). As far as we know, Ref~\cite{Lewandowski:2004yr} is the only attempt to compute the holographic flows beyond the classical level. In that prescription \cite{Lewandowski:2004yr}, integrating out the sliver of geometry between two radial positions does lead to a holographic RG equation resembling Polchinski's exact RG equation~\cite{Polchinski:1983gv}.}

\item{Last but not the least, it is desirable to work out systematically corrections for non-RG quantities such as the coefficients of the double-trace operator in the OPE of the single-trace operators. Then one is equipped with the complete set of the dynamical CFT data to construct four-point functions at $\calo(1/N^2)$.}
\end{itemize}

\acknowledgments{We are grateful to Matt Reece for intensive discussions at all stages of the project and comments on the manuscript. We thank Liam Fitzpatrick for explaining his works and many useful communications. We also thank Richard Brower, Igor Klebanov and Matt Strassler for useful discussions. J. F is supported by the DOE grant DE-FG02-91ER40671.}

\bibliography{c}
\bibliographystyle{jhep}

\appendix

\section{Normalization of the double-trace operators in Eq. 3.18}
It is most convenient to work out the normalizations of double-trace operators in Eq. 3.18 in the position space. There we have the single-particle Green function $G(z_0,w_0;\x,\0)$
\beqs
&&G(z_0,w_0;\x,\0)= c_\Delta U_\Delta(\xi), \nonumber \\
&&c_\Delta=\frac{\Gamma(\Delta)}{2\pi^{d/2}\Gamma(\Delta-d/2+1)} \,, \nonumber \\
&&U_\Delta(\xi)=\xi^{\Delta/2}F(\Delta,\frac{d}{2},\Delta+1-\frac{d}{2},\xi).
\eeqs
In the last line, $\xi \equiv e^{-2\sigma(z_0,w_0;\x,\0)}$ where $\sigma(z_0,w_0;\x, \0)$ is the geodesic distance between the two points $(z_0, \x)$ and $(w_0, \0)$. One property of the functions $U_\Delta(\xi)$ that will be relevant shortly is that they satisfy the orthogonality condition
\beqs
\oint \frac{d\xi}{2\pi i} \frac{(1-\xi)^d}{\xi^{1+d/2}} U_{d-\alpha} (\xi) U_\beta(\xi) =\delta_{\alpha\beta},
\label{eq: ortho}
\eeqs 
where the contour is centered at $\xi=0$.

The two-point function $\langle \phi^2(z_0, \x) \phi^2(w_0, \0) \rangle$ could either be obtained from the Wick contractions
\beq
\langle \phi^2(z_0, \x) \phi^2(w_0, \0) \rangle = 2 G_\Delta(z_0,w_0;\x, \0)^2,
\eeq
or from the sum of all the two-particle propagators with appropriate normalizations
\beq
\langle \phi^2(z_0, \x) \phi^2(w_0, \0) \rangle = \sum_n \frac{1}{N_n^2}G_{\Delta_n}(z_0,w_0;\x, \0).
\eeq
Applying the orthogonal relation, we obtain
\beq
 \frac{1}{N_n^2} = 2\frac{c_\Delta^2}{c_{\Delta_n}}\oint \frac{d\xi}{2\pi i} \frac{(1-\xi)^d}{\xi^{1+d/2}}  U_\Delta(\xi)^2 U_{d-\Delta_n}(\xi).
\eeq 
In $d=2$ and $d=4$, $U_\Delta(\xi)$ turn out to be elementary functions and the integration is simplified greatly. The final results are presented in Eq.~\ref{eq: norm}.

\end{document}